\documentclass[prc,nofootinbib,showpacs,twocolumn]{revtex4}

\usepackage{graphicx}
\usepackage[usenames,dvipsnames]{color}
\usepackage{amsmath,amssymb,bbold}
\usepackage{hyperref}

\begin{document}

\title{Reaction-diffusion equation for quark-hadron transition
in heavy-ion collisions}

\author{Partha Bagchi}
\email{partha@iopb.res.in}
\author{Arpan Das}
\email{arpan@iopb.res.in}
\author{Srikumar Sengupta}
\email{srikumar@iopb.res.in}
\author{Ajit M. Srivastava}
\email{ajit@iopb.res.in}
\affiliation{Institute of Physics, Bhubaneswar, Odisha, India 751005}

\begin{abstract}
Reaction-diffusion equations with suitable boundary conditions 
have special propagating solutions which very closely resemble
the moving interfaces in a first order transition. We show that
the dynamics of chiral order parameter for chiral symmetry breaking
transition in heavy-ion collisions, with dissipative dynamics, is 
governed by one such  equation, specifically, the Newell-Whitehead 
equation. Further, required
boundary conditions are automatically satisfied due to the geometry
of the collision. The chiral transition is, therefore, completed
by a propagating interface, exactly as for a first order transition,
even though the transition actually is a crossover for relativistic
heavy-ion collisions. Same thing also
happens when we consider the initial confinement-deconfinement transition 
with Polyakov loop order parameter. The resulting equation, again
with dissipative dynamics, can then be identified with the reaction-diffusion
equation known as the Fitzhugh-Nagumo equation which is used in population 
genetics.  We discuss the implications of these
results for heavy-ion collisions. We also discuss possible extensions 
for the case of early universe.
\end{abstract}

\maketitle

\section{INTRODUCTION}

Dynamics of quark hadron transition is one of the most important issues
in relativistic heavy-ion collisions, as well as in the universe.
Earlier it used to be believed that the quark-hadron transition is
first order even at low chemical potential (as in the early universe).
This gave a very important proposal by Witten \cite{witten} about
the possibility of formation of quark nuggets due to the concentration of
quarks by moving phase boundaries at the quark-hadron transition.
Dynamics of first order transition also had important implications
for heavy-ion collisions \cite{heavyion}. Subsequently, lattice
results showed that the quark-hadron transition is not first order,
rather it is most likely a cross-over for low chemical potential.
This cross-over is believed to govern the dynamics of transition
in relativistic heavy-ion collisions at high energies. (Though for
lower energies, transition may become first order when the baryonic
chemical potential is sufficiently large.)

For the dynamics of the phase transition, the most important
difference between a first order transition and a cross-over
(or a continuous transition) is the presence of a phase boundary for
the former case which separates the two phases. The transition for
a first order case is completed by nucleation of bubbles which expand.
The moving bubble walls (phase boundaries) lead to physical phenomena,
such as non-trivial scattering of quarks, local heating, specific
types of fluctuations, etc., which are qualitatively different from
the case of cross-over or a continuous transition.

   It turns out that the presence of moving interfaces is more generic,
and not necessarily restricted to the case of first order transitions. Such 
situations routinely arise in the study of so called 
{\it reaction-diffusion equations} \cite{reacdif,solns}, 
typically studied in the context of
biological systems, e.g. population genetics, and chemical systems. 
Typical solution of such equations, with appropriate boundary conditions,
consists of a traveling front with well defined profile, quite like
the profile of the interface in a first order transition case. Importance
of these traveling fronts in the context of high energy physics has been
recognized relatively recently in several works \cite{hep}. In the present
work we demonstrate such solutions for chiral phase transition and
confinement-deconfinement (C-D) transition in QCD even when the underlying
transition is a cross-over or a continuous transition. The only difference 
between the field equations in relativistic field theory case and the
reaction-diffusion case is the absence of second order time derivative
in the latter case. Thus, correspondence between the two cases is
easily established in the presence of strong dissipation term leading to
a dominant first order time derivative term. Further, we will show
that the required boundary conditions for the existence of such a {\it 
traveling front} naturally arise in the context of relativistic heavy-ion 
collision experiments (RHICE). For the case of the universe also it may 
happen in special situations as we will discuss below. 

We would like
to clarify that the propagating front we consider here is like a 
phase boundary (as in a first order transition), and has nothing
to do with hydrodynamic flow. So, the front will still move,
converting one phase (say chirally symmetric phase) to the other
phase (chiral symmetry broken phase) even if the plasma is completely
static. However, the QGP produced in RHICE undergoes hydrodynamic
expansion for which one can use either Bjorken's boost invariant
scaling model for longitudinal expansion, or a 3-d expansion
expected to be applicable at late stages of plasma evolution.
These are incorporated simply by using appropriate metric
for the field equations and they lead to a term of the form
${{\dot \phi} \over \tau}$ where the time derivative is with respect
to the proper time $\tau$. It remains to be explored how this new
type of phase transition dynamics can be incorporated in studies
of full relativistic hydrodynamical evolution of the plasma.

There is a wide veracity of reaction-diffusion equations, see, e.g.
ref.\cite{reacdif,solns}. We will discuss specific equations which can be
identified with the  field equations for the chiral transition and the C-D
transition in QCD in strong dissipation limit. Subsequently we will discuss
different situations in the context of RHICE, with realistic dissipation, 
and show that propagating front solutions of these equations still persist, 
making the dynamics of the relevant transitions effectively like a first 
order phase transition.

\section{REACTION-DIFFUSION  EQUATION FOR CHIRAL TRANSITION}

From the form of these reaction-diffusion equations it will be
clear that such traveling front solutions will exist when the underlying
potential allows for non-zero  order parameter in the vacuum state,
along with a local maximum of the potential \cite{reacdif,solns}. 
The corresponding values of the order parameter provide the required 
boundary conditions for the propagating front solution.
First we consider the case of spontaneous chiral symmetry breaking
transition for the two flavor case with the chiral order parameter
being the O(4) field $\phi = (\sigma, {\vec \pi})$. We will consider
the situation in the context of RHICE, and
study the transition from chiral symmetry (approximately) restored phase to
the chiral symmetry broken phase when the partonic system hadronizes
during the evolution of QGP. For the plasma evolution at this stage,
we will consider longitudinal expansion as well as spherical expansion
(which may be more appropriate for late stages of hadronization).
The field equations are \cite{bynsk}:

$${\ddot \phi} - \bigtriangledown^2 \phi + \eta {\dot \phi} =
-4\lambda \phi^3 + m(T)^2\phi + H$$
\begin{equation}
m(T)^2 = {m_\sigma^2 \over 2} (1 - {T^2 \over T_c^2})
\end{equation}

Here, $\phi$ is taken along the $\sigma$ direction. $T$ is the
temperature and time derivatives are w.r.t the proper time $\tau$.
We have characterized the dissipation term here in terms of $\eta$
which is not a constant for expanding plasma. For Bjorken 1-D scaling
case $\eta = 1/\tau$, while for the spherical expansion $\eta = 3/\tau$.
We again mention that we are only using field equations in the background
of expanding plasma where expansion is incorporated by using time
dependent background metric. It will be interesting to study this
phase transition dynamics in full relativistic hydrodynamical evolution of
the plasma. 
The values of different parameters are taken as \cite{bynsk},
$\lambda = 4.5, m_\sigma = 600$ MeV, and $T_c = 200$ MeV. The coefficient 
of explicit symmetry breaking term $H = (120 MeV)^3$. In the chiral
limit there is a second order phase transition with the critical
temperature $T_c$. In view of the explicit symmetry breaking term
we will take $T = 150$ MeV which allows for the presence of the
central maximum in the effective potential.

 To establish exact correspondence with the reaction-diffusion
equation, we will first neglect explicit breaking of chiral symmetry
(i.e. $H = 0$ in Eq.(1)). Let us also consider the extreme dissipative
case of large value of $\eta$ which is time independent, and neglect the
${\ddot \phi}$ term. For the resulting equation, we rescale the
variables as, $x \rightarrow m(T) x$, $\tau \rightarrow {m(T)^2 \over
\eta} \tau$, and $\phi \rightarrow 2{\sqrt{\lambda} \over m(T)} \phi$.
The resulting equation is,

\begin{equation}
{\dot \phi} =  \bigtriangledown^2 \phi -\phi^3 + \phi
\end{equation}

This equation, in one dimension with
$\bigtriangledown^2 \phi = d^2\phi/dx^2$, is exactly 
the same as the reaction-diffusion equation known as the Newell-Whitehead 
equation \cite{reacdif,solns}. The term $d^2\phi/dx^2$ is the diffusion term
while the other term on the right hand side of Eq.(2) is the so called {\it
reaction term} (representing reaction of members of biological species
for the biological systems). 
       
This equation, in one dimension with
$\bigtriangledown^2 \phi = d^2\phi/dx^2$, is exactly
the same as the reaction-diffusion equation known as the Newell-Whitehead
equation \cite{reacdif}. The term $d^2\phi/dx^2$ is the diffusion term
while the other term on the right hand side of Eq.(2) is the so called {\it
reaction term} (representing reaction of members of biological species
for the biological systems).  We will briefly recall  the analytical 
traveling front solutions for the Newell-Whitehead equation for the 
present case. Subsequently we will study the solutions numerically which 
will help us in obtaining traveling front solutions while retaining the 
${\ddot \phi}$ (and with $\eta$ being time dependent as for the 
expanding plasma).
        
\section{PROPAGATING FRONT SOLUTIONS FOR CHIRAL TRANSITION}

\subsection{Analytical Solution}

Non-trivial traveling front solutions for 
the Newell-Whitehead equation arise when suitable boundary conditions
are imposed, namely $\phi = 0$ and 1 at $x \rightarrow \pm \infty$.
The analytical solution with these boundary conditions has the form,

\begin{equation}
\phi(z) = [1+exp(z/\sqrt 2)]^{-1}
\end{equation}

where $z = x-v\tau$. $v$ is the 
velocity of the front \cite{solns} and has the value $v = 3/\sqrt 2$
for this solution. The reaction-diffusion equations typically have 
several solutions, each with different propagation speeds \cite{solns}. 
For example, Eq.(2) also has a static solution of the form $tanh(z)$. 
Such a solution can have very important implications for 
RHICE as well as for cosmology. We will later briefly comment on it. 
For now we continue with the above analytical solution (Eq.(3)). In Fig.1a 
we show the propagation of this front. For this we have solved Eq.(2)
using leapfrog algorithm of second order accuracy. We have also added
the second order time derivative for numerical solution of Eq.(2) 
for numerical stability and also for comparison with solutions
of full Eq.(1). The requirement of dissipation dominated dynamics
is fulfilled by keeping the $\eta$ coefficient of $\dot \phi$ term large,
with constant $\eta = 10$. 
This introduces a simple scaling of velocity by a factor $1/\eta$.
Solid curve in Fig.1a shows initial profile of the analytical solution
in Eq.(3).  Plots at subsequent times show the 
propagation of the front. The velocity of the front is numerically obtained 
directly by determining the velocity of the front (specifically a 
particular point on the front, say $\phi = 0.5 \phi_{max}$).
We find $v = 0.21$ for $\eta = 10$ in complete agreement with the
scaled velocity $v = {1 \over \eta}{3 \over \sqrt{2}}$. 

For the case of chiral symmetry breaking transition in relativistic 
heavy-ion collisions, the  boundary conditions required for the
solution of Eq.(2) naturally arise
due to radial profile of energy density of the plasma. The center
of plasma represents highest temperature $T_{cntr}$ which smoothly decreases
to values less than the chiral transition temperature $T_c$ in the 
outer regions of the plasma. Thus, at any time when $T_{cntr} > T_c$,
the chiral field will take chirally (approximately) symmetric value
(which will be zero when $H = 0$) at center $r = 0$ and will take
symmetry broken value at large distances.   
We will take such an initial profile, and evolve it when $T_{cntr}$ also 
reduces to a temperature $T_0$ below $T_c$. For simplicity, we will 
assume $T_0$ to be uniform over the range of the profile of $\phi$
with $\phi$ in the center of the plasma having a value $\phi_0$ 
(corresponding to the central maximum of the potential at $T = T_0$).
$\phi$ in outer regions of the plasma will take vacuum expectation value
$\xi$ for $T = T_0$. With such boundary conditions, the analytical
solution in Eq.(3), written  in terms of the original (unscaled) field 
and parameters of Eq.(1), takes the form, 

\begin{equation}
\phi(z) = \xi [1+exp({m(T) \over \sqrt{2}} (x-v\tau))]^{-1},
\end{equation}

where $\xi = \frac{m(T)}{2\sqrt{\lambda}}$ is the vacuum expectation value 
of $\phi$ (for $H = 0$) and the velocity becomes $v = {3m(T) \over \eta 
\sqrt{2}}$. Interestingly, the profile of this analytical solution
is similar to the Wood-Saxon form.
From the energy density profile expected for colliding heavy nuclei, 
a Wood-Saxon type profile for the field is rather natural. 

It turns out that the form of the traveling front, and its evolution,
is essentially unaffected even if we take non-zero value of $H$ in
Eq.(1). Thus, we calculate the numerical profile of the front using
non-zero value of  $H = (120 MeV)^3$. This changes the boundary conditions
for $\phi(z)$. For parameter choice in Eq.(1), the vacuum expectation value 
of $\phi$ is found to be $\xi$ = 75.18 MeV while the central 
maximum of the potential is shifted to $\phi = \phi_0 =$ -25.93 MeV. 
The above analytical solution in Eq.(4), suitably modified for 
these changed boundary conditions, becomes,

\begin{equation}
\phi(z) = - {(\xi - \phi_0) \over A_0} [1 + 
exp({m(T)(|z|-R_0) \over \sqrt{2}})]^{-1} + \xi
\end{equation}

where the normalization factor  $A_0 = [1 + exp({-m(T)R_0 \over 
\sqrt{2}}]^{-1}$. Here we use $|z|$ in order to have symmetric 
front on both sides of the plasma for the present 1-d case.
$R_0$ represents width of central part of the plasma.

\subsection{Numerical Solution}

  We calculate numerical solutions for the full Eq.(1), retaining
the ${\ddot \phi}$ term. To correspond to the analytical solution, 
we first consider a large, constant, value of $\eta = (1/0.14) 
fm^{-1}$ in Eq.(1) (even though chiral transition occurs late at
$\tau \simeq$ 4-5 fm), and a uniform fixed $T = 150$ MeV. Note
that this does not represent realistic QGP evolution in RHICE.
We first study equations with constant (and uniform) $T$ {\it only}
to show exact correspondence with traditional reaction-diffusion equations.
We will see that the resulting propagating front is exactly the same
as discussed in literature for reaction-diffusion equations. Subsequently
we relax this assumption of constant $T$ and study proper time dependence
of $T$ for expanding QGP. We still retain the assumption of uniform 
temperature for studying front propagation as with spatially varying
$T$ the effective potential also has to vary spatially and correspondence
with reaction-diffusion equation becomes remote.  We consider 1-d 
case as  suitable for the Newell-Whitehead equation. This will be 
applicable when  size of traveling front is large, so a planar 
approximation can be used. Fig.1b gives the numerical profiles of the  
front $\phi(z)$ at different times (with initial time taken as 4 fm). 
The front starts at a distance of about 10 fm from the center and
moves inwards converting the central region to chiral symmetry broken 
phase. Note, as $T_0$ everywhere has a value corresponding to the
symmetry broken phase, one would have expected rapid roll down of the
field to $\phi = \xi$ everywhere in time of order 1 fm. In complete
contrast to this, we see that phase conversion here happens slowly,
by the motion of well defined interface, just like for the case of a
first order phase transition. As for Fig.1a, we obtain $v$ here also
directly from the traveling front. We find $v$ ranging from 0.41 to 
0.35 in close agreement with the expected value $v = {3m(T) \over \eta 
\sqrt{2}}$ (= 0.42). 
By finding propagating solutions for different values of $\eta$ we have
verified (for both cases, Fig.1a, and Fig.1b) that the velocity
of the front exactly scales as $1/\eta$.

\begin{figure}[!htp]
\begin{center}
\includegraphics[width=0.45\textwidth]{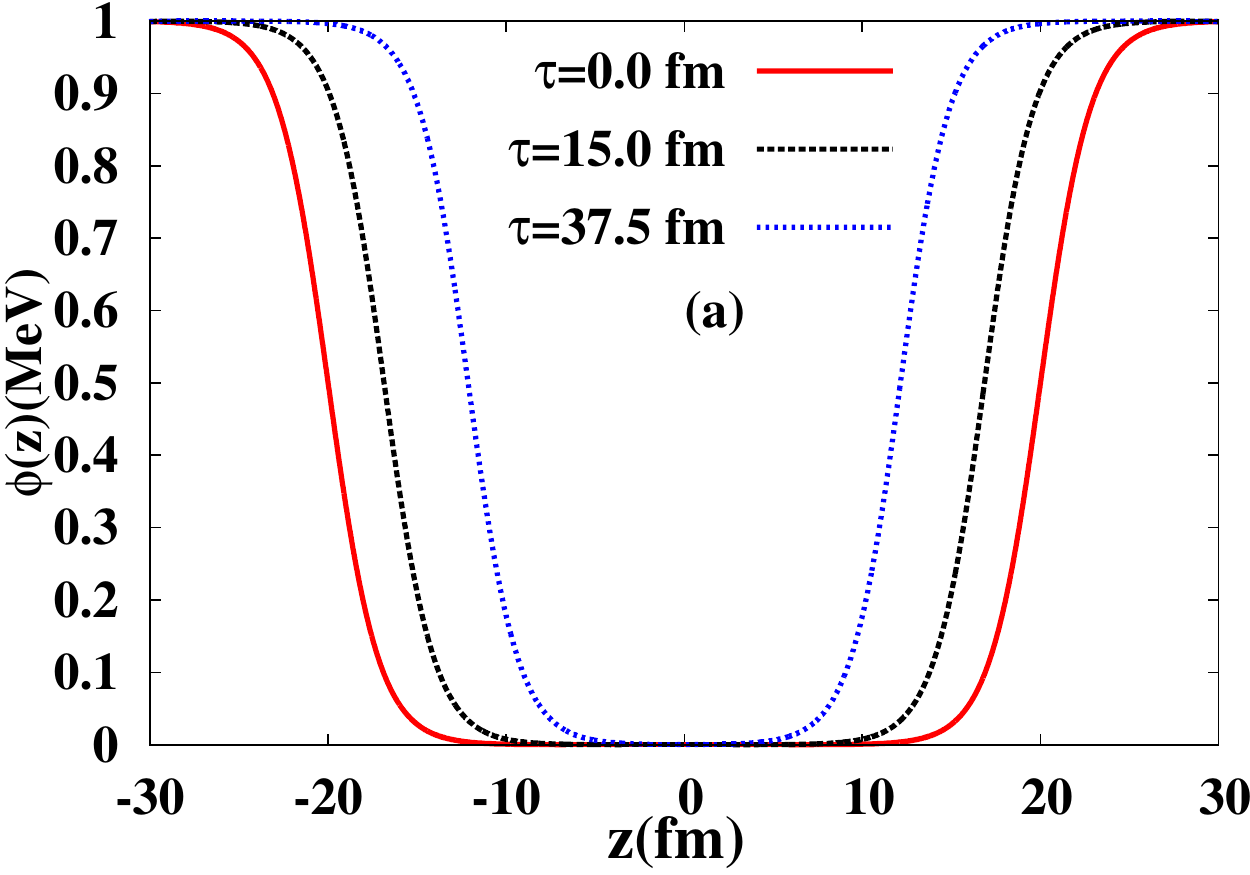}
\includegraphics[width=0.45\textwidth]{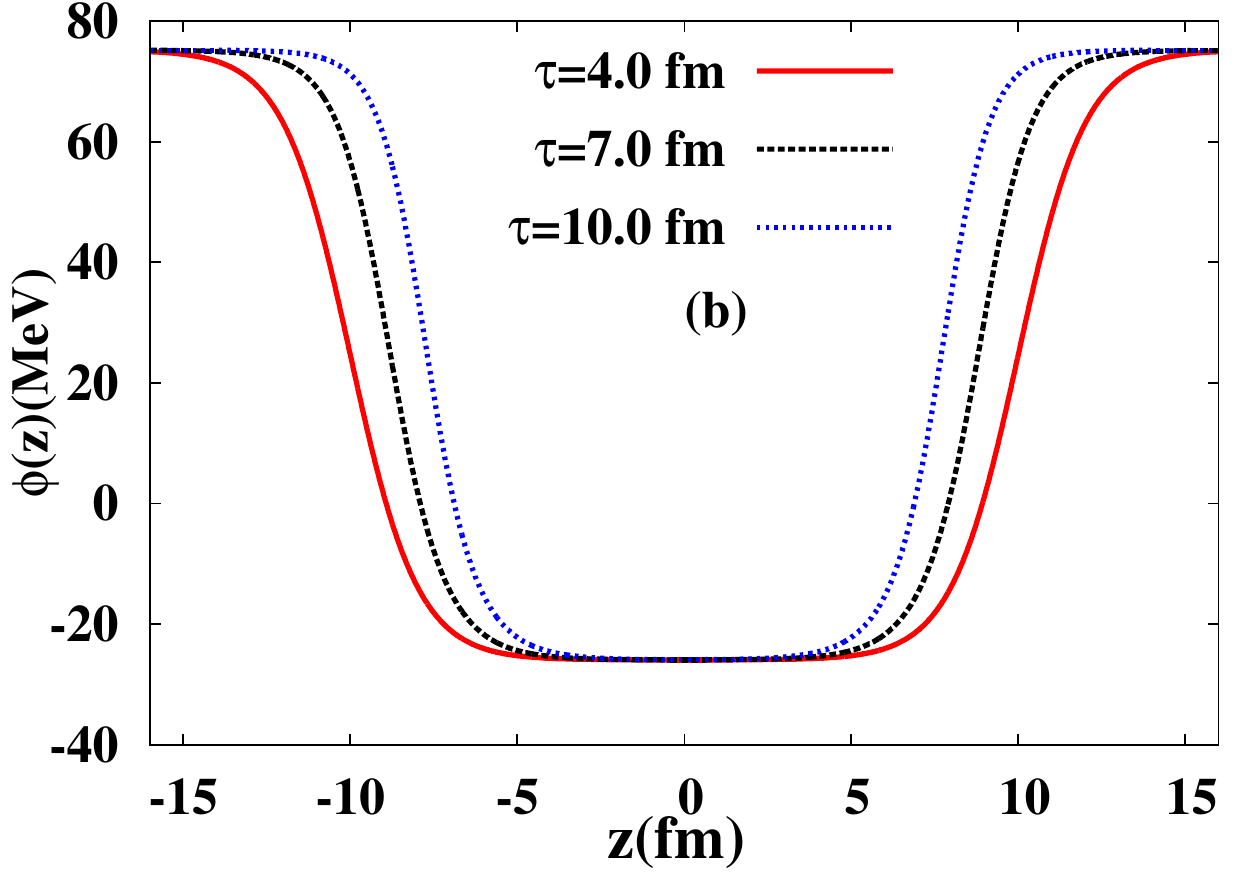}
\caption{(a) Plots of the numerical solution for the traveling front 
of Eq.2 (as discussed in the text) at different times.
(b) shows the numerical 
solution for profile in Eq.(5) with non-zero $H$.}
\label{fig1}
\end{center}
\end{figure}

The propagating front solutions we obtain are very robust and almost
independent of the initial profile of the front taken. To show this,
we show evolution of a profile consisting of linear segments (with
correct boundary conditions) in Fig.2a. We see that this also develops 
into a well defined propagating front as shown in Fig.1. 
Next, we consider realistic values of time dependent $\eta = 1/\tau$ so
that second time derivative term becomes important, with
initial value of $\tau = \tau_0 = 4.0$ fm, as appropriate for the 
chiral transition. We also take  $T(\tau) = T_0 (\tau_0/\tau)^{1/3}$ in
accordance with Bjorken's scaling solution for the longitudinally
expanding plasma. Fig.2b shows the traveling front solution for this case at 
different values  of proper time $\tau$. The only minor difference from
plots in Fig.1 is seen at somewhat later stages, with a little rise at the 
boundary of the front. The central value of $\phi$ changes in accordance 
with time dependent $T$.
It is clear that with the presence of this {\it front} structure most of
the results for first order transition, such as non-trivial quark 
scattering, well defined phase separated regions, fluctuations etc.  
will become applicable.

\begin{figure}[!htp]
\begin{center}
\includegraphics[width=0.45\textwidth]{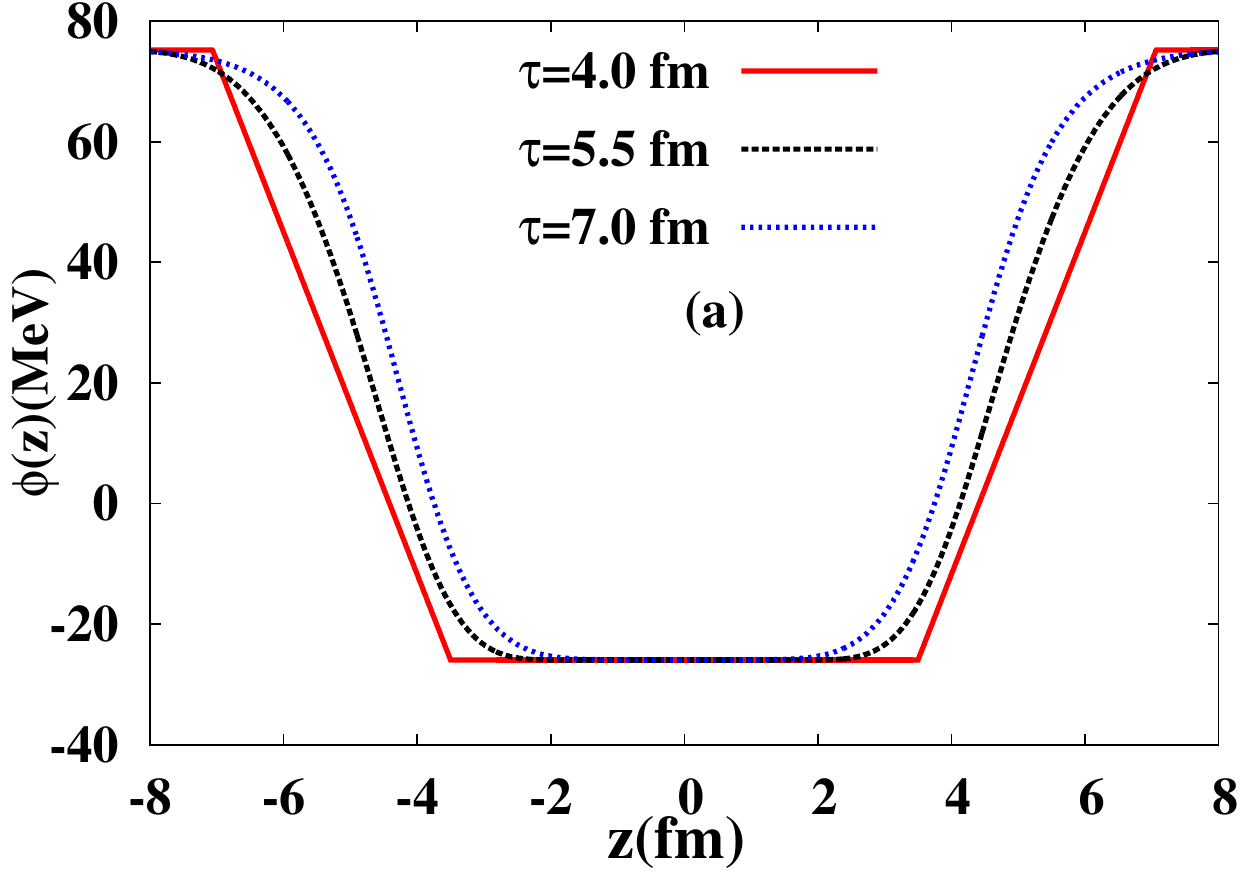}
\includegraphics[width=0.45\textwidth]{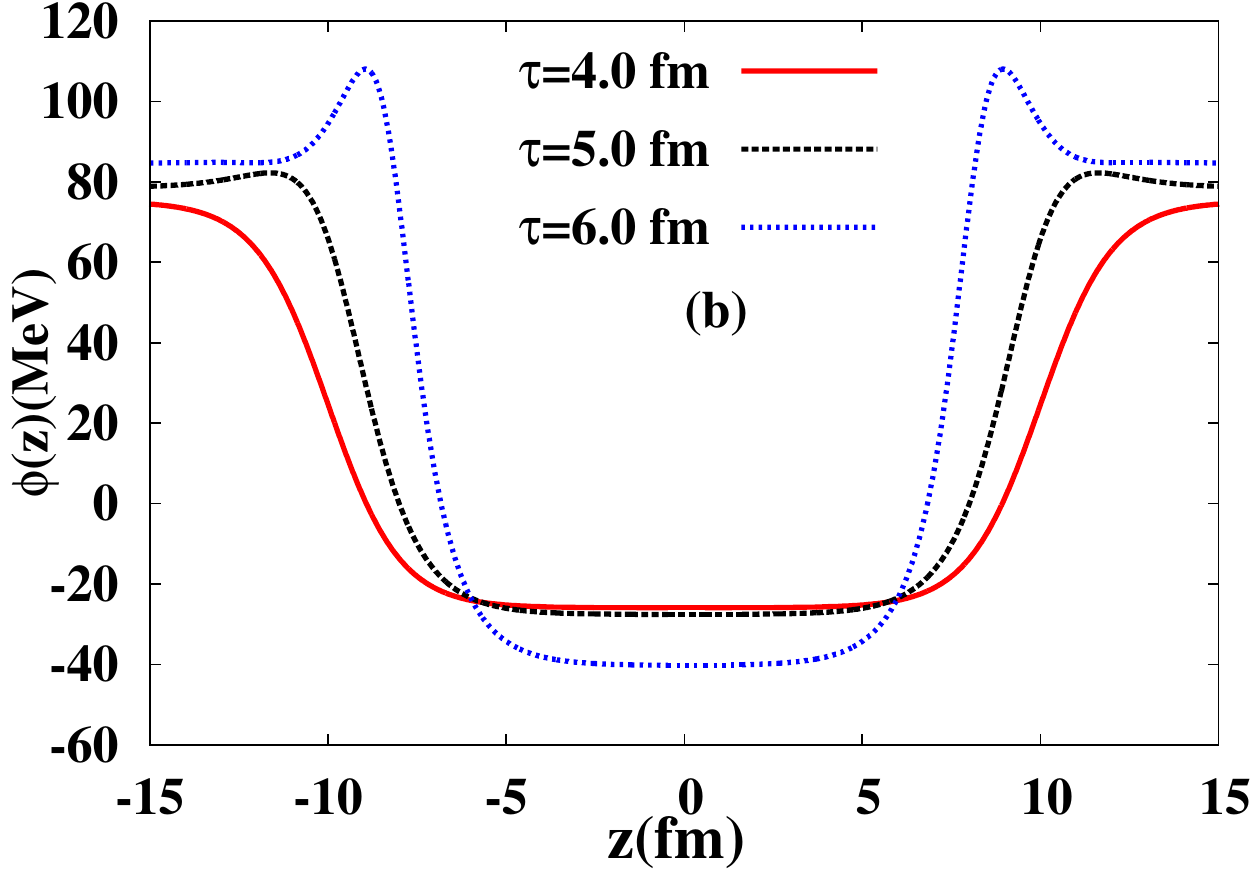}
\caption{(a) Initial profile consisting of linear segments also rapidly
evolves into a well defined propagating front as in (a).
(b) Solutions of Eq.(1), with realistic values of time dependent 
$\eta$ and time dependent $T$.} 
\label{fig2}
\end{center}
\end{figure}

For Bjorken's 1-d longitudinally expanding plasma, one should consider 
transverse motion of the front. Neglecting transverse expansion of the 
plasma, one should use Eq.(1) with cylindrical coordinates, or spherical
expansion for the late stages of plasma evolution.
Eq. (1) for these cases will become
 
\begin{equation}
{\ddot \phi} - {d^2\phi \over dr^2} - {d-1 \over r}{d\phi \over dr}  
+ \eta {\dot \phi} = 
-4\lambda \phi^3 + m(T)^2\phi + H
\end{equation}
   
where $d$ = 2 for Bjorken's 1-d longitudinal expansion, and $d$ = 3
for spherical expansion. We have obtained front solutions for both
these cases here for the chiral symmetry case as well as later for 
the Z(3) case of C-D transition. The resulting solutions are very similar 
to those obtained for 1-d case (as in Figs.1,2). Hence we do not show
those plots.

\section{REACTION-DIFFUSION EQUATION FOR CONFINEMENT-DECONFINEMENT 
TRANSITION}

 We now consider the case of confinement-deconfinement transition during
early thermalization stage. For RHICE, for very early stages, Bjorken
scaling with longitudinally expanding plasma is a very good approximation.
The dissipation term is very strong during very early stages which helps
in making a direct correspondence with reaction-diffusion equations. As
the traveling wave solutions exist in the symmetry broken phase we will
consider the case of initial confinement-deconfinement transition using
the Polyakov loop order parameter with center symmetry Z(3) spontaneously
broken in the high temperature QGP phase. This early stage
actually represents a non-equilibrium stage, with system thermalizing
to a maximum temperature $T_0$ in a time scale $\tau_0$. The
thermalization time scale $\tau_0$ can be as short as about 0.14 fm
(at LHC). Elliptic flow measurements indicate an upper bound of about
1 fm on $\tau_0$. For the dynamics of the order parameter $l(x)$ which
is the expectation value of the Polyakov loop, we use the
following effective Lagrangian density \cite{zn}.

\begin{equation}
L={N\over g^2} |{\partial_\mu l}|^2{T^2}- V(l)
\end{equation}

Where the effective potential $V(\it l)$ for the Polyakov loop, in case
of pure gauge theory is given as

\begin{equation}
 V(l)=({-b_2\over2} |l|^2- {b_3\over 6}( l^3+( l^ \ast)^
 3)+\frac{1}{4}(|l|^2)^2){b_4{T^4}}
\end{equation}

At low temperature where $\it{l}$ = 0, the potential has only one
minimum. As temperature becomes higher than $T_c$ the Polyakov loop
develops a non vanishing  vacuum expectation value $l_0$, and
the $l^3 + l^{*3}$ term above leads to $Z(3)$ generate vacua. Now in 
the deconfined phase, for a small range of temperature above $T_c$, 
the $\it{l} =$ 0 extremum becomes the local
minimum (false vacuum) and a potential barrier exist between the local
minimum and global minimum (true vacuum) of the potential. As we
are interested in showing the existence of traveling front solutions
in the absence of any first order transition, we will consider value
of temperature $T$ to be sufficiently large so that there is no such barrier
present. For the parameter values we use, this requires $T > 280$ MeV, and
we take $T = 500$ MeV. The values of various coefficients in Eq.(8) are
the same as used in our previous works
\cite{gupta,gupta2} (including discussions about explicit symmetry breaking
strength $b_1$) and we do not repeat that discussion here. (With those
values of parameters, the transition temperature is taken to be
$T_c = $ 182 MeV.)

For simplicity we neglect the effect of dynamical
quarks which lead to explicit breaking of Z(3) symmetry, and hence a
linear term in $l$ in $V(l)$ above \cite{gupta2}. This can be taken care
of in a similar manner as the explicit symmetry breaking term $H$ for
the chiral symmetry case in Eq.(1). Similarly, as our interest is not
in the Z(3) structure of the vacuum, we will take $l$ to be real. We again
first neglect the second order time derivative (for large dissipation
case).  The variables are scaled as, $x \rightarrow gT 
\sqrt{b_4 \over 2N} x$, and $\tau \rightarrow 
{b_4 g^2 T^2 \over 2 \eta N} \tau$. With that, the field equation for 
(real) $l(x)$ can be written as follows (for the sake of uniformity,
we will denote $l(x)$ as $\phi(x)$ in the following),

\begin{equation}
{\dot \phi} =  \bigtriangledown^2 \phi +\phi (b_2 + b_3 \phi - \phi^2)
\end{equation}

 The final equation in this case is again a reaction-diffusion
equation known as the Fitzhugh-Nagumo equation which is used in population
genetics \cite{reacdif,solns}. Thus we again expect well defined traveling
wave solutions for appropriate boundary conditions.

\section{PROPAGATING FRONT SOLUTIONS FOR CONFINEMENT-DECONFINEMENT TRANSITION}

Required boundary conditions for the propagating front solution for
Eq.(9) again naturally arise in RHICE, during
early stages. As the system thermalizes, one expects first the center
of the plasma to reach a temperature $T > T_c$ where $T_c$ is the C-D
transition temperature. The temperature in (somewhat) outer regions remains
below $T_c$ initially. This leads to a profile of $l(x)$ where $l = 0$ in
the other regions while $l = 1$ at the center of the plasma. Subsequently
even these regions, somewhat away from the center, also achieve $T > T_c$.
With these boundary conditions, we solve equations for $l(x)$ with a uniform
temperature $T$ with initial value = 500 MeV. Eq.(9) above was derived in
the large dissipation limit to identify it with the Fitzhugh-Nagumo equation
which guarantees the existence of a traveling wave solution for $l(x)$.
With that assurance, we will now solve the full field equations for $l(x)$ 
(i.e. for $\phi(x)$) including the  second time derivative term. The 
dissipation term is naturally large initially in this case due to $1/\tau$
factor with $\tau_0$ being very small. Again, to show direct correspondence
with reaction-diffusion equation, we will take a very large, fixed,
value of $\eta = 1/\tau_0^\prime$ with $\tau_0^\prime = 0.01$ fm, and will 
keep temperature to have fixed value $T = T_0 = $ 500 MeV.

Plots in Fig.3a show the well defined traveling wave solution
at different values of $\tau$ starting from initial time taken as
$\tau_0 = 0.14$ fm. The initial profile is taken to have similar form 
as in Fig.1, suitably modified for the boundary conditions appropriate for
the present case.  We next consider realistic value of time dependent 
$\eta = 1/\tau$ with initial value of $\tau = \tau_0 = 0.14$ fm, and
take $T(\tau) = T_0 (\tau_0/\tau)^{1/3}$
as appropriate for the Bjorken 1-d scaling solution. Resulting
evolution of $\phi(z)$ is shown in Fig.3b. Though $\phi$ shows
some oscillations, still it shows a reasonably well defined 
propagating front. It is possible that $\eta$ may not decrease
as fast as $1/\tau$ due to presence of other sources of dissipation.
In that case resulting solution will be closer to that in Fig.3a.
 
\begin{figure}[!htp]
\begin{center}
\includegraphics[width=0.45\textwidth]{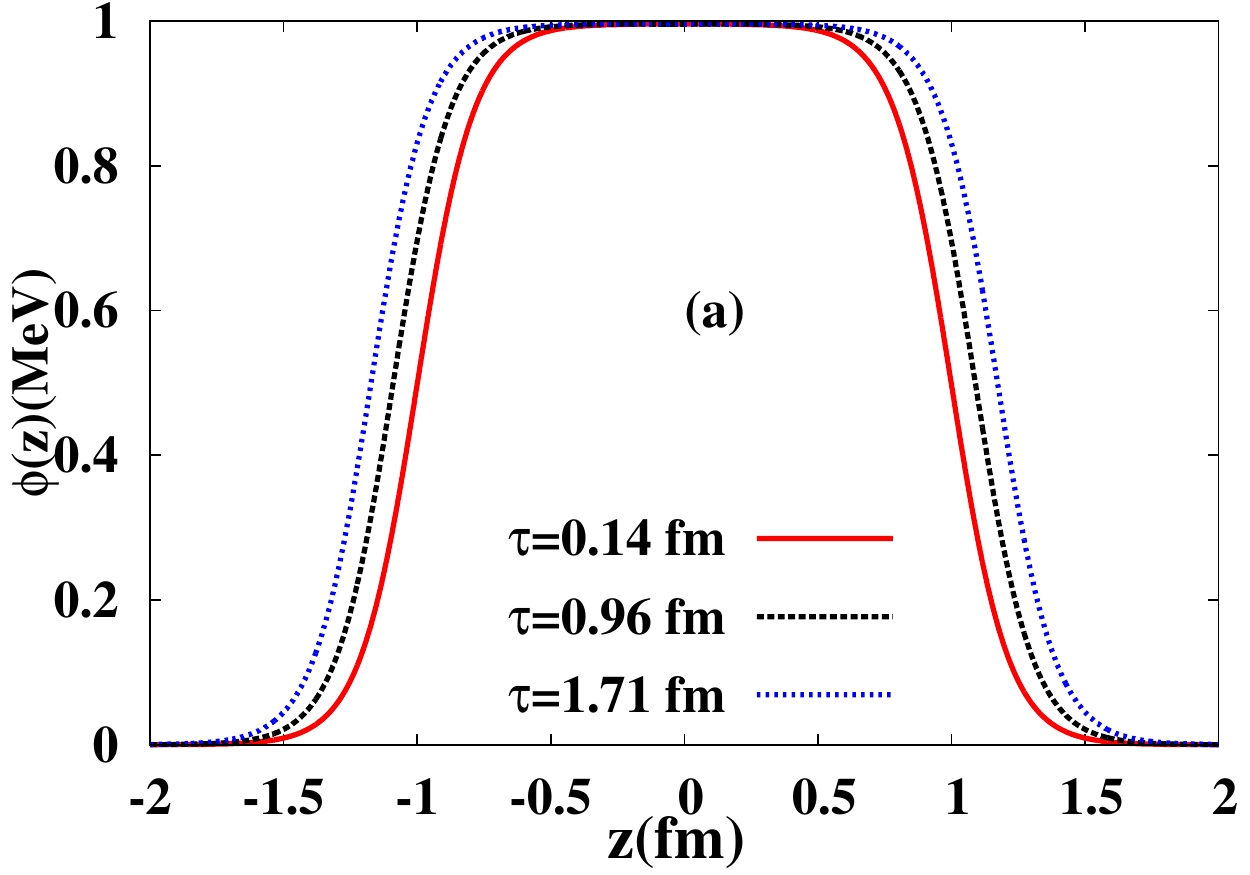}
\includegraphics[width=0.45\textwidth]{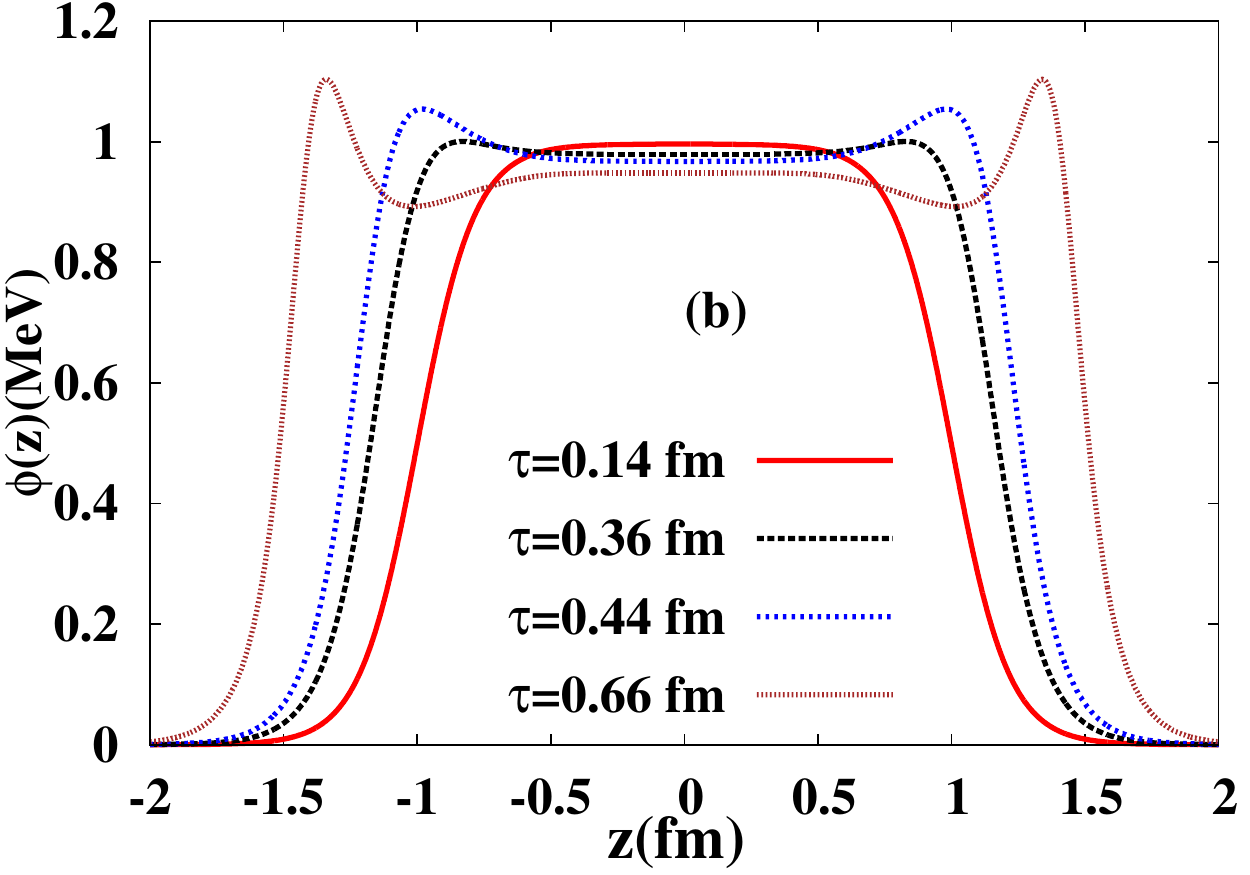}
\caption{(a) Traveling wave solution for the Polyakov
loop order parameter at different times for very large dissipation 
case with constant $\eta$ = 1/(0.01fm). (b) Solution for realistic $\eta = 
1/\tau$ and with time dependent $T$.}
\label{fig3}
\end{center}
\end{figure}

\section{RELATION BETWEEN CHIRAL TRANSITION AND DECONFINEMENT TRANSITION} 

We now address an important issue regarding the difference in
the behavior of order parameter evolutions for the chiral symmetry
case and for the Polyakov loop for the C-D transition case.
Note that we discussed traveling wave solution for the chiral
symmetry case during late stages of the evolution, even though
presence of large dissipation case was harder to justify for
such late stages ($1/\tau$ factor being relatively smaller).
Similarly, for the early thermalization stage, we only discussed
the case of C-D transition with the Polyakov loop, and did not discuss
the chiral symmetry case. The reason is that the traveling front
solutions via the reaction-diffusion equation approach arises only
when the effective potential has specific shape, for example, that
corresponding to spontaneous symmetry breaking. Thus, spontaneous breaking
of Z(3) symmetry during early thermalization leads to well defined
traveling front for the Polyakov loop. But during this stage, chiral
symmetry is restored. There is absolutely no possibility of
finding any traveling wave solution for the symmetry restored
effective potential for the chiral field $\phi$ (as one can simply
check analytically as well as numerically). Similarly, while a
traveling front exists for the chiral field during chiral symmetry
breaking transition during late stages of the plasma, there is
no such solution for the Polyakov loop order parameter at that stage
as the Z(3) symmetry actually gets restored. This raises serious
concerns with the conventional idea that the two transitions,
namely the chiral transition and the C-D transition, are somehow related
(or, are the same). In our calculations we clearly find regions
separated by the traveling front of one order parameter whereas
no such phase separation is expected from the other order
parameter. Thus, we will conclude that our results support the claims
of several groups that the chiral transition and the C-D transition
are indeed separate, and independent, transitions. In fact, our
results provide clearly phase separated regions (by the traveling front)
which are, e.g. chirally symmetric but in confined phase, or in
chiral symmetry broken phase but in the deconfined phase.

\section{CONCLUSIONS}

 We conclude by emphasizing that the techniques of reaction-diffusion
equation have been used here to show existence of well defined
traveling front solutions, which are very similar to phase boundaries
for a first order transition case, even though the relevant QCD transitions
here are of second order, or a cross-over.  This allows the very exciting
possibility of using earlier results valid for a first order
transition case, such as formation of strangelets, baryon concentration, 
fluctuations etc. for RHICE.  Our results show that the transition 
proceeds by slow moving front, and may take several fm time to complete,
leading to long lasting mixed phase stage. This will affect calculations
of various signals of QGP for RHICE, e.g. production of
thermal photons and di-leptons, $J/\psi$ suppression, and especially
elliptic flow which develops mostly during early stages.
As we mentioned above, the reaction-diffusion equations have other 
solutions also with different propagation speeds. For
example, Eq.(2) also has a static solution of the form $tanh(z)$.
If the initial density profile in RHICE leads to such a profile of
$\phi(z)$ the transition will become stagnant. This will have
important implications for RHICE. It is clearly of great importance to 
see if these results can be applied to the case of early Universe
(for example, solutions with different speeds, especially the one with
zero speed). The required
initial boundary condition of $\phi = 0,1$ for $x \rightarrow \pm \infty$
looks difficult to justify for the Universe (due to absence of a
temperature profile as for RHICE). However, one should remember that there
are always density (and hence temperature) fluctuations present in the
Universe (most likely of inflationary  origin). Though these are
very tiny (one part in 10$^5$), but if we consider these fluctuations
when the temperature of the Universe is very close to the transition
temperature $T_c$ then there can easily be regions of space where the
symmetry is restored, while other neighboring regions will have
symmetry broken phase. 

This will lead to the required boundary
conditions for the traveling wave fronts as discussed here. Small 
magnitude of temperature fluctuations will imply small difference 
in the magnitude of $\Phi$ at the two boundary points, 
subsequently leading to small effects (like scattering of quarks). 
However, typical wavelength of these fluctuations will be comparable
to the Hubble size, naturally leading to wavefront propagation over
such large scales. Thus, even with small quark scattering etc. one
may be able to get large concentration of baryons via Witten's mechanism
of nugget formation.

  One caveat in this scenario is that chiral symmetry (as well as
Z(3) symmetry) are also explicitly broken. This makes the above mentioned 
scenario difficult to implement for the quark-hadron transition when 
temperature fluctuations have very small magnitude. There may be other 
possibilities for quark-hadron transition which we intend to explore in 
a future work. However, such a mechanism will certainly apply for
electroweak phase transition where there is no explicit symmetry
breaking involved. There also one will get traveling front solutions
arising from inflationary fluctuations irrespective of the order, or the 
strength of phase transition. It will be interesting to investigate how 
such front solutions can affect the physics of post electroweak transition
physics, in particular sphaleron mediated baryogenesis etc.

\section*{Acknowledgment}
We thank Debashis Ghoshal for introducing us to the Fisher
equation. We are very grateful to Trilochan Bagarti for
discussions on reaction-diffusion equations and for very
useful references. We also thank Shreyansh S. Dave and 
Biswanath Layek for useful discussions.


\begin{thebibliography}{99}

\bibitem{witten} E. Witten, Phys. Rev. {\bf D 30}, 272 (1984).

\bibitem{heavyion} S. Digal and A.M. Srivastava, Phys. Rev.
Lett. {\bf 80}, 1841 (1998).

\bibitem{reacdif} A.G. Nikitin, T.A. Barannyk, Central 
European Journal of Mathematics 2(5), 840 (2004);
B. Bradshaw-Hajek, {\it Reaction-diffusion 
equations for population genetics}, PhD thesis,  School  
of Mathematics  and Applied Statistics,  University of 
Wollongong,  2004, http://ro.uow.edu.au/theses/201;

\bibitem{solns} {\it Progress in Nonlinear Differential Equations
and Their Applications}, Editor H. Brezis, Springer Basel AG,
Switzerland (2004).

\bibitem{hep} R. Peschanski, Phys. Rev. {\bf D}, 81,
054014 (2010); S. Munier and R. Peschanski, Phys. Rev. Lett. 
{\bf 91}, 232001 (2003); D. Ghoshal, JHEP, {\bf 1112}, 015 (2011);
D. Ghoshal and P. Patcharamaneepakorn, JHEP, {\bf 1403}, 015 (2014).

\bibitem{bynsk} D. Boyanovsky, H. J. de Vega, R. Holman, and S. 
Prem Kumar, Phys. Rev. {\bf D 56}, 3929 (1997).

\bibitem{zn} O. Scavenius, A. Dumitru, and J. T. Lenaghan,
Phys. Rev. {\bf C66}, 034903 (2002); R.D. Pisarski, Phys. Rev. 
{\bf D62}, 111501R (2000).

\bibitem{gupta} U. S. Gupta, R. K. Mohapatra, A. M. srivastava and
V. K. Tiwari, Phys. Rev.{\bf D82}, 074020 (2010).

\bibitem{gupta2} U. S. Gupta, R. K. Mohapatra, A. M. srivastava and 
V. K. Tiwari, Phys. Rev. {\bf D 86} 125016 (2012).

\end{thebibliography}
\end{document}